\documentclass[aps,prd,showpacs,floats,nofootinbib,superscriptaddress,preprint,tightenlines]{revtex4}

\usepackage[dvips]{graphicx}
\usepackage[mathscr]{eucal}
\usepackage[latin1]{inputenc}
\usepackage{amsmath}
\usepackage{amsfonts}
\usepackage{amssymb}
\usepackage{pictex}
\usepackage{epsfig}
\usepackage{colordvi}
\usepackage{color}

\def\be{\nopagebreak[3]\begin{equation}}
\def\ee{\end{equation}}
\def\ba{\nopagebreak[3]\begin{eqnarray}}
\def\ea{\end{eqnarray}}
\def\bas{\nopagebreak[3]\begin{eqnarray*}}
\def\eas{\end{eqnarray*}}

\def\d{{\rm d}}

\newcommand{\teta}{\rlap{\lower2ex\hbox{$\,\tilde{}$}}\eta{}}

\newcommand{\bi}{\begin{itemize}}
\newcommand{\ei}{\end{itemize}}

\newcommand{\mb}[1]{\mathbb{#1}}

\def\h{\hat }

\def\lp{{\ell}_{\rm Pl}}

\def\bb{{\tt b}}

\def\co{\sqrt{12 \pi G}}
\def\la{\langle}
\def\ra{\rangle}

\newcommand{\f}{\frac}

\usepackage{enumerate}

\usepackage{colordvi}
\newcounter{mnotecount}[section]

\newcommand{\comment}[1]{}

\def\f{\frac}

\def\t{\tilde}
\def\chip{\chi_+}
\def\chim{\chi_-}

\begin{document}
\preprint{\vbox{\baselineskip=12pt \rightline{IGC-11/5-2}
}}

\title{On the Semiclassical Limit of Loop Quantum Cosmology}
\author{Alejandro Corichi}\email{corichi@matmor.unam.mx}
\affiliation{Centro de Ciencias Matem\'aticas,
Universidad Nacional Aut\'onoma de
M\'exico, UNAM-Campus Morelia, A. Postal 61-3, Morelia, Michoac\'an 58090,
Mexico}
\affiliation{Center for Fundamental Theory, Institute for Gravitation \& the Cosmos,
Pennsylvania State University, University Park
PA 16802, USA}
\author{Edison Montoya}
\email{edison@ifm.umich.mx}
 \affiliation{Instituto de F\'{\i}sica y
Matem\'aticas,  Universidad Michoacana de San Nicol\'as de
Hidalgo, Morelia, Michoac\'an, Mexico}
\affiliation{Centro de Ciencias Matem\'aticas,
Universidad Nacional Aut\'onoma de M\'exico,
UNAM-Campus Morelia, A. Postal 61-3, Morelia, Michoac\'an 58090,
Mexico}

\begin{abstract}
We consider a $k$=0 Friedman-Robertson-Walker (FRW) 
model within loop quantum cosmology (LQC) and explore the
issue of its semiclassical limit.  The model is exactly solvable and allows us to construct analytical (Gaussian) coherent-state solutions for each point on the space of classical states. We propose physical criteria that select from these coherent states, those that display semiclassical behavior, and study their properties in the deep Planck regime. Furthermore, we consider generalized  {\it squeezed} states and compare them to the Gaussian states. 
The issue of semiclassicality preservation across the bounce is studied and shown to be generic for all the states considered. Finally, we comment on some implications these results have, depending on the topology of the spatial slice. In particular we consider the issue of the recovery, within our class of states, of a {\it scaling symmetry} present in the classical description of the system when the spatial topology is non-compact. 
\end{abstract}

\pacs{04.60.Pp, 04.60.Ds, 04.60.Nc}
\maketitle

\section{Introduction}

Given a quantum theory describing a physical system, a pressing question is the existence of its 
semiclassical limit; particularly so if the quantum theory was the result of the quantization of a
classical system. In this case, one would like to know whether it is possible to recover the original system in some
limit. In particular, one is interested in exploring the existence of 
semiclassical states that exhibit the dynamical properties of the original classical system. It is only when this step can be completed that one can assign the status of physically viable to the formalism in question.
Assuming that one has solved the question of existence of the limit, 
uniqueness of such semiclassical states is an issue that one might be concerned with. That is, can we uniquely specify a state with the aid of some preferred semiclassicality conditions?
Is there some freedom in finding a state given a classical configuration? How can we parametrize this freedom?
These and similar questions have motivated the study of semiclassical states, including the case of constrained systems
(See for instance \cite{semiclassical-coherent} and references therein). 

The system that we shall consider here is a
quantum cosmological model found by applying {\it loop quantization} methods \cite{lqg} to a flat FRW model coupled to a massless scalar field \cite{lqc,aps0,aps2,closed,open,polish,vp,slqc,cs:unique,geometric}. In this case, the quantum theory that describes the evolution of the (homogeneous and isotropic) universe has the property that the classical big bang, represented by a singularity of the theory, is
replaced by a `quantum bounce'. The quantum evolution does not stop at the would be singularity and defines a spacetime much larger than the classical general relativity (GR) equations suggest. This simple system is of particular interest since
it has been shown to be described by a solvable model \cite{slqc}, with which it has been possible to prove that the bounce is generic and some relevant quantities such as energy density and the expansion parameter of observers
are absolutely bounded \cite{slqc,geometric}. Furthermore, it has been shown that a state that satisfies very general semiclassicality conditions on one side of the bounce, has to preserve those conditions on the other side \cite{cs:prl,kam:paw}.

The issue we want to address in this manuscript is the construction of semiclassical states given a classical configuration of the geometric and matter degrees of freedom for our model. That is, can we find semiclassical states
given any classical configuration? 
Another issue that is intertwined with the question at hand is the so called `classical regime' of any cosmological model. Let us be more precise. Since every classical trajectory in the general relativity (GR) dynamics has two very different regimes we have to be very precise to define what we mean by the classical regime. On the one hand, an expanding solution, say, has a regime of very low energy density, and `large volume' where we expect GR to be a very good description. On the other hand, when evolving back in time, the solution approaches a regime of very high density, end eventually reaches the singularity where the density diverges. One expects the regime near to the singularity not to be well described by GR, but instead belong to the quantum realm. The transition from one regime to the other is expected to be, on purely dimensional arguments, near the Planck scale.
This is precisely what we see in the LQC dynamics. When the energy density approaches the Planck scale, the dynamics departs from GR, a quantum repulsive force dominates and the universe bounces. Thus, it is only in the low density, GR regime that we expect to be able to impose semiclassicality conditions in order to select the states we are looking for.  Even for those states that behave semiclassicaly in the low density regime,
one might expect that in the Planck regime, the state might loose its coherence and behave `wildly'.

The departure of the LQC dynamics from GR seems to be very well described by a so-called ``effective dynamics"
constructed from the quantum theory \cite{victor,wkb}. This `classical' description captures the most important quantum effects coming from loop quantum geometry, and approaches the GR dynamics in the classical, low density regime. It is natural then to try to test the validity of the effective description into the deep quantum regime. Finally, another question that one would like to address is that of the preservation of semiclassicality across the bounce. If one uses the low density, GR regime, to specify a semiclassical state, what is then its behavior on the other side of the bounce? It is still semiclassical? Does it loose coherence and becomes highly quantum?

The purpose of this paper is to address all these questions, using the solvable model of loop quantum cosmology introduced in \cite{slqc}, and further studied in \cite{cs:prl}. In a sense, the solvable model is the ideal arena, since one can find exact coherent state solutions for which all these questions can be met with precise analytical answers. This analysis also allows us to put in concrete, analytic and quantitative terms some of the results that were derived in \cite{cs:prl,kam:paw}, regarding the preservation of semiclassicality across the bounce for general semiclassical states. As we shall see, the class of states we consider, build out of
coherent (and squeezed) states are enough to specify semiclassical states with the desired properties and to put the effective theory to the test. As we shall see in detail, there are some semiclassicality conditions to be imposed that imply that not all trajectories on the classical phase space can be recovered by a semiclassical state, signaling the existence of a semiclassical regime and a purely quantum regime within the space of quantum states. 
Another objective of this manuscript is to discuss the physical motivations that lead to the definition of semiclassical states an discuss their main properties. Details of the model and of the derivations of some of the results used here have appeared elsewhere \cite{CM1}. Furthermore, here we put forward a discussion regarding the issue of {\it cell independence} present in the $k$=0 model, and its role in the construction of coherent semiclassical states.
Let us start by specifying the model we are considering.

\section{The System: A $k$=0 FRW model}

The model we are considering, namely a $k$=0
FRW cosmology is given by the line element,
\be
\d s^2= -\d t^2 + a^2(t) \d\Sigma^2
\ee
where $\d\Sigma^2$ corresponds to a flat metric on the three-surface $\Sigma$.
The matter content is a massless scalar field
$\phi$, satisfying the Klein-Gordon equation. Phase space $\Gamma$ consists of
gravitational degrees of freedom: {$(a,p_a)$} and
matter degrees of freedom: {$(\phi, p_\phi)$}
In terms of the coordinates
$(a,p_a,\phi,p_\phi)$ for the kinematical phase space $\Gamma$ of the
theory, all the  ``dynamics" is captured in the Hamiltonian constraint
\be
{\cal C}:=-\frac{3}{8}\;\frac{p_a^2}{|a|}+8\pi
G\,\frac{p_\phi^2}{2|a|^3}\approx 0 
\ee
The diffeomorphism constraint is satisfied, and its corresponding gauge is fixed. 
The equations of motion are such that
$\dot{\phi}=8\pi\,G\frac{p_\phi}{|a|^3}$ has the same sign
as $p_\phi$ and never vanishes. Thus $\phi$ can be used as an internal (or relational)
time variable.
The homogeneous, isotropic, sector of the Ashtekar formulation
can be expressed, for flat FRW as,
\be
A_a^i=c\,V_0^{-(1/3)}\,{}^{\rm o}\omega^i_a
\quad{\rm and}\quad E^a_i=p\, V_0^{-(2/3)}\,\sqrt{q_0}\,{}^{\rm o}e_i^a\, ,
\end{equation}
where both connection and triads are written in terms of fiducial
quantities. The fundamental Poisson bracket between the degrees of freedom $(c,p)$ 
is given by, $\{c,p\}=\frac{8\pi G\gamma}{3}$, 
with $\gamma$ the Barbero-Immirzi parameter of LQG. 
It is convenient to define the following new coordinates:
${{\tt b}:=\frac{c}{|p|^{1/2}}}$ and ${{V}:=p^{3/2}}$.
The new coordinate ${\tt b}$ is then equal to ${\tt b}=\frac{c}{a}$. On the constraint surface, that is, on classical solution to the equations of motion it becomes, ${\tt b}=\frac{\gamma\dot{a}}{a}=\gamma\,H_{\rm Hubble}$.
The Poisson bracket now becomes,
\be
\{{\tt b},{V}\}=4\pi G\gamma\, .
\ee
In these new coordinates, the constraint is: $
{\cal C}:= -\frac{6}{\gamma^2}\,V\,{\tt b}^2+8\pi G\,\frac{p^2_\phi}{V}=0$.
A convenient choice is to select the lapse function as $N=V$ such that $\t{\cal C}:=N {\cal C}$ becomes
\be
\tilde{\cal C}:= -\frac{6}{\gamma^2}\,V^2\,{\tt b}^2+8\pi G\,{p^2_\phi}=0\, .
\ee
Thus, the phase space {$\Gamma \leftrightarrow (V,{\tt b},\phi,p_{\phi})$}, is 4-dimensional.
The constraint defines a 3-dimensional pre-symplectic space {$\bar{\Gamma}$}. The
{\it physical phase space} {$\hat{\Gamma}=\bar{\Gamma}/\sim$} is therefore 2-dimensional.
A natural question that arises has to do with the choice of Dirac observables to parametrize it and therefore
represent the true degrees of freedom. That is, what coordinates on $\hat{\Gamma}$ are most convenient to define semiclassical states peaked around them? How many parameters do we need?
The first obvious choice is to select $p_\phi$, since it is a Dirac observable.
Since we need another observable to characterize the physical phase space, what remains to be seen is whether  we can use $V$ or $\tt b$ as the other coordinate, or some other possibility.

There are several strategies to define semiclassical states. We could classify them, in very broad terms as in \cite{BBC,semiclassical-coherent} by,

\noindent
i) Kinematical semiclassical states (KSS), peaked on a point of the
constrained surface $\bar{\Gamma}$.

\noindent
ii) Start with a KSS and project to a dynamical semiclassical state using, for instance the {\it group averaging method}.

\noindent
iii) Define a dynamical semiclassical state peaked on an appropriate point of the physical phase space.

\noindent
iv) If the system admits a deparametrization, define a `peaked initial state' and evolve to generate 
a physical state.

While identifying what is the best approach is not the purpose of this letter, here we shall adopt approach iv) in what follows, since our system admits a deparametrization in terms of internal time $\phi$. 
Therefore, our strategy is
to construct suitable initial data to generate physical states. Furthermore, in the remaining of the manuscript we shall impose physically motivated conditions to select those states that behave semiclassically. 
As we shall see below, the 
class of initial states we shall consider is of the {\it coherent state} type.

\section{Solvable Loop Quantum Cosmology (SLQC)}

Let us recall the basic features of the solvable model
in LQC, as developed in \cite{slqc} (See also \cite{CM1}).
The starting point is the quantum constraint,
\be
\left(\sin(\lambda \bb)\hat{\nu}\sin(\lambda \bb)\hat{\nu} + K\hat{P}_\phi^2\right)\cdot\chi=0
\ee
with $K$ a constant.
Let us now take functions $\chi({\tt b},\phi)$. In this case the would-be operator $\hat{{\tt b}}$, that would act by multiplication, does not exist on the kinematical Hilbert space. Thus, we have to define an operator to regulate it.  The LQC choice is to have $\frac{1}{\lambda}\,\sin(\lambda{\tt b})\cdot\,\chi({\tt b},\phi)$, and the operator $\hat{V}$ now acts by derivation: $
\hat{V}\cdot\chi:=-i4\pi\ell_{\rm Pl}^2\gamma\,\frac{\partial{}}{\partial {\tt b}}\,\chi$.
The variable {\tt b} turns out to be periodic, taking values in $(0,\pi/\lambda)$.
The quantum constraint of SLQC is given by,
\be
\f{\partial^2}{\partial \phi^2}\cdot\chi(\bb,\phi)=12\pi G\,
\left(\frac{\sin(\lambda\bb)}{\lambda}\;\frac{\partial}{\partial\,
\bb} \right)^2
\cdot \chi(\bb,\phi)\label{const1}
\ee
with $\bb\in (0,\pi/\lambda)$ and $\chi(\bb,\phi)=-\chi(\pi/\lambda-\bb,\phi)$.
We can now define a  new coordinate $x$ by
\be
x = \f{1}{\co} \, \ln \left|\tan\left(\f{\lambda
\bb}{2}\right)\right| ~. 
\ee
Then the basic constraint equation (\ref{const1}) translates to
\be
\partial_\phi^2 \, \chi(x,\phi) = \partial_x^2 \, \chi(x,\phi) ~.
\ee
A general solution { $\chi(x,\phi)$} to the above equation can be
decomposed in to left moving and right moving components:
\be\label{chi_eq} { \chi } = {\chi_+ (\phi + x)} +
{{\chi_-(\phi - x)}} :=
{ \chi_+(x_+)} +
{ \chi_-(x_-)} 
\ee
such that ${ \chip}$ and ${ \chim}$ satisfy $
i \, \partial_\phi \, { \chip} =  i \, \partial_x \, {\chip}$ and  
$i \, \partial_\phi \, {\chim} =  - i \, \partial_x \, {\chim}$.

\noindent
Furthermore, we are interested in {\it positive frequency solutions} to the Klein Gordon equation.
We can then write a `Schr\"odinger  equation':
\be
- i\,\partial_\phi \chi(x,\phi) = \sqrt{-\partial^2_x}\;\chi(x,\phi)\, ,
\ee
with which we can evolve an initial state to produce a physical state.
The general form of the physical states is:
${\chi(x,\phi)}:=\f{1}{\sqrt{2}}\,(F(x_+)-F(x_-))$, for $F$ an arbitrary positive
momentum solution. This means that
\be
{ \chi(x,\phi)} = 
{ \int_0^\infty\,\d k
\;\tilde{F}(k)\,e^{-ik(\phi+x)} }-
{ \int_0^\infty\,\d k \;\tilde{F}(k)\,e^{-ik(\phi-x)} }\, .
\ee
This form implements the required antisymmetry:
$\chi(x,\phi)=-\chi(-x,\phi)$. 
The physical inner product on solutions to the quantum constraint is given by,
\be
(\chi_1,\chi_2)_{\rm phy}=i\int\d x\,[ (\partial_x\bar{F}_1(x_+))\, F_2(x_+)-
(\partial_x\bar{F}_1(x_-))\,F_2(x_-)]\, .
\ee
The expectation value for $\hat{V}|_{\phi}$, is given
by:
\be
\langle\hat{V}\rangle_\phi=V_+\,e^{-\alpha\,\phi}
+V_-\,e^{\alpha\,\phi} \, ,
\ee
where $\alpha= \sqrt{12\pi\,G}$. There is a bounce for all states in the physical Hilbert space.
The bounce time $\phi_{\rm b}$ is then,
$\phi_{\rm b}:=\frac{1}{2\,\alpha}\ln\left(\frac{V_+}{V_-}\right)$.
The coefficients $V_\pm$ are given by \cite{slqc},
\be V_{\pm}=\frac{4\pi \gamma \lp^2
\lambda}{\alpha}\int\d x \left|\frac{\d F}{\d
x}\right|^2\,e^{\pm\alpha\, x} \, .
\ee
Thus, they are strictly positive and can be computed from the `initial state'.
The time dependence is then,
$\langle\hat{V}\rangle_\phi=V_{\rm b}\cosh\left[\alpha(\phi-\phi_{\rm b})\right]$, 
where the minimum value of the volume is given by $V_{\rm b}=2\,\sqrt{V_-\,V_+}$.
In the same fashion, one can define the operator $\hat{x}|_{\phi}$ for which the expectation value can be computed (see \cite{CM1} for details).

\section{Gaussian States}

Let us consider physical states with initial
states given {\it generalized Gaussian states}, defined 
by functions $\tilde{F}(k)$ of the form:
\be \label{coh_state}
\tilde{F}(k)=\left\{
\begin{array}{rl}
k^n e^{-(k-k_0)^2 /\sigma^2}e^{-ip_0 k},& \mbox{\rm for }\; k>0 ,\\[1ex]
0,& \mbox{\rm for }\; k\le 0 ,
\end{array}
\right.
\ee
with $\sigma>0, k_0>0$, $ n=0,1,2,...$.
That is, we are choosing Gaussian states centered around the point
$(p_0,k_0)$, with `dispersion' given by $\sigma$. The system possesses
two phase space {\it physical} degrees of freedom, and from our experience with, say, a harmonic oscillator,
coherent states labeled by two phase space coordinates and with an extra parameter ($\sigma$) are over-complete and enough to describe the semiclassical sector of the theory. As we shall see below, the states under consideration above will also be enough to describe semiclassical states for the cosmological system.
The question is:
In what coordinates is this identification made?
Namely, what is the point in physical phase space where the state is peaked?
We expect {$k_0$} to be the point on the variable `canonically conjugate to {$x$}', and $p_0$ to represent the point where it is peaked in $x$. 
A careful analysis of the (gauge fixed) physical phase space for the effective theory shows that, in
$(x,p_\phi)$ coordinates the symplectic structure has the form \cite{CM1}:
$\hat{\Omega}=\d p_\phi\wedge\d x$.
This means that in the physical phase space, as defined by the effective theory, one expects $p_\phi$ to be canonically conjugate to $x$. One can also expect that
the parameter $k_0$ be related to the expectation value for $p_\phi$. As we shall now see, these expectations
are realized in the quantum theory. These considerations also suggest that we can select the classical trajectory
to be approximated by the quantum state, by specifying the pair $(\bar{x},\bar{p}_\phi)$ at, say, $\phi=0$, and then choosing the parameters $(p_0,k_0)$ appropriately so that the state is sharply peaked around the classical point.

In the case of pure Gaussian states, namely for $n=0$, we have the expectation value on the physical state space to be,
\be \langle \hat p_\phi \rangle_{\rm phy} = \hbar k_0 \left[1 +\f{\sigma^2}{4 k_0^2} \right]  \, .\label{exp-val-p}
\ee
If we fix a tolerance parameter $\delta$ that bounds how far from the classical value $\bar{p}_\phi$ the expectation value
of the operators are: $|\la\hat{p}_\phi\ra- \bar{p}_\phi| < \delta$. This condition then imposes some interval on 
which $k_0$ can take values. For $\bar{p}_\phi\gg\hbar\sigma$ and $\bar{p}_\phi\gg\delta$ we have that
$k_0$ can take values in the interval: 
$\bar{p}_\phi/\hbar - \delta/\hbar +(\delta^2-\hbar^2\sigma^2)/4\hbar \bar{p}_\phi \leq k_0 \leq
\bar{p}_\phi/\hbar + \delta/\hbar +(\delta^2-\hbar^2\sigma^2)/4\hbar \bar{p}_\phi$.\\
Note that if the above conditions are satisfied, the interval is, to a good approximation, of size $2\delta/\hbar$ and centered around the point $\bar{p}_\phi/\hbar$.
The dispersion takes the form
\be
(\Delta \hat p_\phi)^2 = \f{\hbar^2\sigma^2}{4}\left[1 -\f{\sigma^2}{4 k_0^2}\right]\, ,
\ee
which is a constant, as expected. Note that, if
$k_0\gg \sigma$, the relative dispersion $(\Delta \hat p_\phi)^2/\langle \hat p_\phi \rangle^2$ is then approximately given by $(\sigma^2/4k_0^2)(1+\sigma^2/4k_0^2)$, which is rather small.
This brings us to our first {\it semiclassicality condition}, namely to impose that $k_0\gg \sigma$. 
This means that the Gaussian is sharply peaked away from the origin
(in $k$, and it is practically zero at $k\leq0$, a condition necessary in our
approximation \cite{CM1}).  Furthermore, this condition naturally yields an expectation value for $\hat{p}_\phi$ close to the classical value $\bar{p}_\phi$, and sharply peaked around it.

We can now consider the operator $\hat{x}$ associated to the coordinate $x$ (for a definition of this symmetric operator and some of its properties, see \cite{CM1}). We have that,
for all values of $n$, the expectation value of $\hat{x}$ reads
\be
 \label{x_0} \la \hat x \ra_\phi= p_0 -\phi\, ,
\ee
which confirms the expectation that the parameter $p_0$ represents
the point $\bar{x}$ in $x$ space, where the Gaussian is peaked (when $\phi=0$). The evolution of $\la \hat x \ra_\phi$ as a function of $\phi$ corresponds precisely to the one defined by the classical {\it effective} dynamics.
The dispersion of the operator is given by \cite{CM1},
\be \label{deltax} (\Delta \hat x)_\phi^2
=\f{1}{\sigma^2}\,\f{\left(1+\f{3\sigma^2}{2k_0^2} \right)}
{\left(1 +\f{3\sigma^2}{4k_0^2} \right)}
 \ee
which is also constant and tells us that the parameter $\sigma$ has the expected
interpretation of providing the inverse of the dispersion for the
observable $x$, when $k_0\gg\sigma$.

The uncertainty relations for the observables
$\hat{p}_\phi$ and $\hat{x}$ (for $n=1$) is then,
\be {(\Delta \hat p_\phi)^2(\Delta \hat x)^2}
= {\f{\hbar^2}{4}}\; \f{\left(1+\f{9\sigma^4}{16k_0^4}+
\f{3\sigma^2}{4k_0^2}-\f{9\sigma^6}{64k_0^6}\right)
\left(1+\f{3\sigma^2}{2k_0^2} \right)}
{\left(1+\f{3\sigma^2}{4k_0^2}\right)^3}\, .
 \ee
We note that, if $k_0\gg\sigma$, then the uncertainty relations
$$
(\Delta \hat p_\phi)(\Delta \hat x) \geq \f{\hbar}{2}\, ,
$$
are very close to being saturated. To summarize, given a point $(\bar{p}_\phi,\bar{\bb})$  on $\hat{\Gamma}$, we
can define a family of coherent states (labeled by $n$), defined by coefficients $(k_0,p_0)$, that are `peaked'
on the point $(\bar{p}_\phi,\bar{\bb})$, provided $k_0=\bar{p}_\phi/\hbar$ and $p_0=\bar{x}:=\frac{1}{\alpha}\ln|\tan(\lambda\bar{\bb}/2)|$. The parameter $\sigma$ is still free, except for the mild condition that it must be small compared to $k_0$. Further conditions need to be imposed to select, from these Gaussian states, those that we shall call semiclassical.

\section{Semiclassicality Conditions}

\begin{figure}[tbh!]
\includegraphics[scale=0.55]{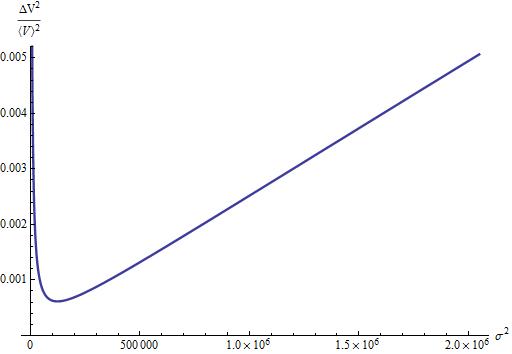}
\caption{The asymptotic relative fluctuation $\Delta_\pm$ is plotted as a function
        of $\sigma^2$ with $k_0=10000$ in Planck units, where
        $\alpha\approx 6.14$. Here we have chosen $n=0$ and $p_0=0$.
        Two features characterize the behavior of the function near
        the minimum $\t\sigma$. The first one is that for $\sigma^2<\t\sigma^2$ the function is
        exponential; the second one is that for $\sigma^2>\t\sigma^2$ the function has a
        polynomial behavior.
}\label{fig:1.1}
\end{figure}

Let us now impose physically motivated conditions on the states to ensure they are semiclassical. 
The most natural choice is to require that the state be sharply peaked in the GR limit. This means that in the {\it large} volume, small density regime, fluctuations should be small. The observable that we still have at our disposal is volume. We have seen that the time evolution (in $\phi$) of the expectation value of the volume has a generic form for {\it all} states, that follows exactly the trajectory defined by the `effective theory'. Thus, in order to select semiclassical states we need to impose criteria on the dispersion of the volume, a quantity 
we know to grow exponentially to the future and past. However, we expect that the relative dispersion be well behaved. Thus our criteria will be to require that
the asymptotic relative dispersion $\Delta_{\pm}:=\lim_{\phi\to\pm\infty}\left( \frac{\Delta\hat{V}}{\la\hat{V}\ra}
\right)_\phi^2$  be small. For that we fix a real number $\varepsilon$, which will measure how semiclassical the state is.
For $n$=0 this condition takes the form \cite{CM1},
\be \Delta_\pm 
=e^{\alpha
^{2}/\sigma^2} \f{\left(1 +\f{3\sigma^2}{4k_0^2}+\f{\alpha
^2}{k_0^2} \right)}{\left(1 +\f{\sigma^2 +\alpha ^2}{4k_0^2}
\right)^2} -1\leq \varepsilon\, .  \label{delta-PM}
 \ee
Note that the appearance of the constant $\alpha=\sqrt{12\pi\,G}$ introduces a scale for $k_0$ and $\sigma$.
Not only do we need that {$\alpha\ll k_0$} and {$\sigma\ll k_0$}, but we are also forced to choose $\sigma$ wisely to make $\Delta_{\pm}$, as defined by Eq.(\ref{delta-PM}), as small as possible (see Fig.~\ref{fig:1.1}).
The condition that the expression (\ref{delta-PM}), as a function of $\sigma$, be
an extrema is a cubical equation and the only physically
interesting solution can be approximated as,
$\tilde{\sigma}^2 \approx 2\,\alpha\, k_0 +\f{2}{7}\, \alpha^2$ \cite{CM1}.
Since we require that $k_0\gg \alpha$, the second
term is very small compared to the first one and we can approximate it by:
$\tilde{\sigma}\approx \sqrt{2\,\alpha\,k_0}$.\\
If we now introduce this value $\t\sigma$ into the expression for the relative
fluctuation, then this can be approximated by
  \be
 \Delta_{\pm}= \left.\lim_{\phi\to\pm\infty}\left[\f{(\Delta V)^2}{\langle V
  \rangle^2}\right]\right|_{\t\sigma}(k_0)
  \approx \f{\alpha}{k_0} \, .
  \label{opt-dispersion}
  \ee
This gives us then a relation between $k_0$ and $\alpha$ depending on $\varepsilon$. That is, the condition on the relative dispersion  (\ref{delta-PM}) will be satisfied provided $k_0>\alpha/\varepsilon$.
We can then see from Fig.\ref{fig:1.1}, that there is only a small range of allowed values for $\sigma$  around
its minimum $\t \sigma$, that yield semiclassical states, specially for $\sigma < \t\sigma$.

To summarize, we have seen that the class of states under consideration, namely generalized Gaussian states are 
enough to completely characterize and describe semiclassical states. In this respect, one is not loosing generality by restricting to these states to describe the semiclassical sector of the theory.

\vskip0.3cm
\noindent
{\it Semiclassicality at the Bounce.} Let us now explore the properties of the semiclassical states at the bounce, that is, in the deep Planck regime. It is easy to see that, for $n$=0, the relative dispersion in volume at the bounce and its asymptotic value are related by,
\be \left. \f{(\Delta V)^2}{\langle V
\rangle^2} \right|_{\rm b}-\f{1}{2}\Delta_{\pm}  =  \f{
e^{-\alpha^2/\sigma^2}\left(1+\f{3\sigma^2}{4k_0^2 }\right)}
{2\left(1+\f{\sigma^2 +\alpha^2}{4k_0^2} \right)^2}-\f{1}{2}\Delta_{\pm} = O(\alpha^2/\sigma^2) +
O(\sigma^2/k_o^2)\, .
\ee
Thus, for semiclassical states satisfying $\alpha\ll\sigma\ll k_0$, the relative dispersion at the bounce is, to a very good approximation,
1/2 of its asymptotic value ($\approx \alpha/2k_0\ll 1$).
This means that the state is not loosing its coherence, even in the deep quantum regime.

\begin{figure}[tbh!]

\begin{center}
\includegraphics[scale=0.6]{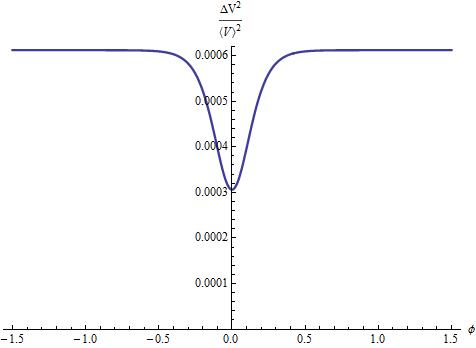}
\caption{The relative fluctuation $(\Delta V)^2/\la\hat{V}\ra^2$
is plotted as a function
        of internal time $\phi$ with $k_0=10000$, in Planck units where
        $\alpha\approx 6.14$. Note that the
        minimum of this quantity is at the bounce (here
        corresponding to $\phi=0$). It is found that the value at
        the bounce is approximately 1/2 of its asymptotic value for
        the generalized Gaussian states. 
}\label{fig:1.2}
\end{center}
\end{figure}


There are several arguments that support the claim that the {\it effective theory} describes the dynamics of semiclassical states in LQC \cite{victor,wkb,ma}. Is that also the case for our semiclassical states?
We can indeed explore this question with our exact coherent states.
For that, we can consider the volume at the bounce $V_{\rm b}$.
The effective theory yields, for each trajectory, a  minimum value for the volume given by,
$ V_{\rm min}=\sqrt{\f{8\pi G \gamma^2
\lambda^2}{6}}\;p_\phi\, .$ 
On the other hand, for $n$=0, the volume at the bounce $V_{\rm b}=2\,\sqrt{V_+V_-}$ is given by \cite{CM1},
\be \label{Vmin} V_{\rm b}=\f{4\pi \gamma \lp^2
\lambda}{\alpha }\;
 e^{\alpha ^2/2\sigma^2}\, k_0
\left[1 +\f{\sigma^2 +\alpha ^2}{4k_0^2} \right]\, .\ee
If we 
recall our previous result that $\langle \hat p_\phi \rangle =
\hbar\, k_0 + O(\sigma^2/k_0^2)$, and 
assume the semiclassically conditions found in previous sections, 
namely that $\sigma  \gg
\alpha$ and $k_0 \gg \sigma$ (as is the case when $\sigma=\t\sigma =
\sqrt{2\alpha k_0}$ and $k_0\gg \alpha$) in Eq. \eqref{Vmin}, we can compare
the last two equations to see that $V_{\rm bounce} \approx \f{4\pi \gamma \lp^2
\lambda}{\alpha }\f{\langle \hat p_\phi\rangle}{\hbar}$. 
Using the value $\alpha=\sqrt{12\pi G}$, we conclude that \be
 V_{\rm bounce} \approx V_{\rm min}\,+{O}(\alpha^2/\sigma^2) +
O(\sigma^2/k_0^2) + O(\alpha^2/k_0^2) \, .
\ee 


We can ask the same question regarding the density at the bounce. The simplest expression for density 
is given by
$\t \rho=\f{\la \h p_\phi\ra^2}{2\la \h V\ra^2}$, a
quantity that was shown to be bounded by $\rho_{\rm crit}=3/(8\pi G\gamma^2\lambda^2)$ \cite{slqc}. It
is straightforward to find this quantity for the $n=0$ states \cite{CM1}, 
\be
\t \rho=\rho_{\rm crit}\;\f{e^{-\alpha^2/\sigma^2}\left(1 +\f{\sigma^2}{4 k_0^2}
\right)^2} { \left(1 +\f{\sigma^2
+\alpha^2}{4k_0^2} \right)^2}\, .
 \label{rho}
\ee 
Just as with volume at the bounce, we see that the more semiclassical the state is, namely, 
the larger $k_0$ is, the better the value given by the effective theory of the density at the bounce.  

So far, we have considered the spatially flat FRW model without specifying the topology of the spatial section $\Sigma$. If the topology of $\Sigma$ is given by a three torus $\mathbb{T}^3$, the interpretation of the variables $V$ and $\bb$ is clear: $V$ is the volume of the universe at `time $\phi$', and $\bb$ is its canonically conjugate variable defined by the (unique) symplectic structure (and closely related to the Hubble parameter in the GR limit). In this case, all our results have a clear interpretation. The semiclassicality conditions imply then that the theory can only have a classical limit if the universe bounces when its volume is much larger than the Planck volume. Therefore, we can not have semiclassical states that bounce near the Planck volume. This is, as we have seen, due to the fact that the relative fluctuations in volume (and $\hat{x}$) are not small in that case.
As previously noted \cite{aps2,closed}, in the case of a spatially closed universe the `Planck scale' is defined by the energy density being close to Planck's density and not by the volume at the bounce, that can take many values. On the other hand, if the topology of the spatial section is $\mathbb{R}^3$ and the universe is open, we have to introduce a fiducial cell to make sense of the Hamiltonian formulation. This introduces an extra freedom that is, however, physically irrelevant. Let us now comment on what our result imply for this scenario.

\section{Rescaling Dependence}

An important issue that results from our analysis is the implementation of
the classical ``gauge freedom" of cell rescaling, present in both the classical and effective descriptions of the system. This symmetry of the classical equations, that can be seen by considering  $p_\phi\rightarrow \ell\,p_\phi$, together with $V\rightarrow \ell\,V$, becomes rather subtle. From our previous results, it might appear that, by restricting ourselves to our class of generalized coherent states, the
classical scaling symmetry can only be recovered in the large $k_0$ limit ($k_0\gg \alpha$). This might at first be surprising, given that in the classical description (and in the effective theory as well) this ``gauge freedom", understood as a mapping between two theories for two different cells, 
can be reinterpreted as a mapping on
each theory defined on a given fiducial cell ${\cal V}$. To be precise, if we have a cell ${\cal V}$ and a second one ${\cal V}'$ obtained from the first one by a constant rescaling by a `rescaling factor' $\ell$, then there is a preferred mapping from the phase space $\Gamma_{\cal V}$ of the theory on cell ${\cal V}$ to the phase space $\Gamma_{{\cal V}'}$ given by $({\bb},V,\phi,p_\phi)_{\cal V}\rightarrow({\bb},\ell V,\phi,\ell p_\phi)_{{\cal V}'}$. Thus, the arbitrariness in the
definition of the fiducial cell, that should yield physically identical descriptions, can be reinterpreted as a  transformation on $\Gamma_{\cal V}$ that should connect two physically indistinguishable configurations. While this is certainly a reasonable assumption, detailed 
considerations suggest that this is not the case. For, a detailed analysis of the phase space transformation shows that it does not correspond to a phase space symmetry. That is, the transformation does not preserve the symplectic structure, even when it connects two solutions. From the viewpoint of constrained systems, it corresponds to a transformation connecting two different gauge orbits, that is not a symplectomorphism in the physical phase space either. In this regard, this symmetry can not be regarded as {\it gauge} in the standard Hamiltonian language.

If this transformation is {\it not} canonical, should we expect its quantum implementation to be unitary? 
In fact, what one actually would like to have is a mapping such that
a state $\psi_{\cal V}$ for the theory on the cell ${\cal V}$ is mapped to a state $\psi'_{{\cal V}'}$ in such a way that the expectation value of $\hat{p}_\phi$ has the appropriate transformation properties:
$\la\hat{p}_\phi\ra_{{\cal V}'}=\ell\,\la\hat{p}_\phi\ra_{{\cal V}}$. If such a map were also to relate the
expectation values of volume by $\la\hat{V}|_\phi\ra_{{\cal V}'}=\ell\,\la\hat{V}|_\phi\ra_{{\cal V}}$, we would have a quantum implementation of `cell rescaling'\footnote{One might further require higher moments to transform appropriately.}. 

Let us now see that this expectation is not realized in this model, for the special class of states under consideration. Recall that our states are labeled by parameters $(k_0,p_0,\sigma)_n$, in such a way that, given
a classical configuration $(\bar{p}_\phi,\bar{x})$ (at time $\phi=0$), there is a preferred choice of
parameters $(k_0,p_0,\sigma)_n$ that select a {\it unique} semiclassical states (for a given $n$).
One could now ask how
would these parameters  `transform' in order to implement the symmetry within our space of solutions.
There is a natural answer for this question. The rescaling symmetry induces a transformation on phase space, and therefore, on the space of labels $(\bar{p}_\phi,\bar{x})$ defining the classical configuration to be described by the quantum states. Thus, there is a {\it unique} implementation of the classical scaling symmetry, on the
space of generalized coherent states under consideration. The question is whether, under such preferred mapping
between states, the expectation values of observables $\hat{p}_\phi$ and $\hat{V}_\phi$ scale appropriately.

Let us now explore this question in further detail.
From Eq.(\ref{exp-val-p}) we can see that, to implement cell rescaling in $\hat{p}_\phi$, the natural mapping between states would be to send a state with labels $(k_0,p_0,\sigma)$ to a state defined by $(k_0'=\ell\, k_0,p_0,\sigma'=\ell\,\sigma)$.
In this way we have $\la\hat{p}_\phi\ra_{{\cal V}'}=\ell\,\la\hat{p}_\phi\ra_{{\cal V}}$. But note that, even when
$\psi_{\cal V}$ is taken to be a semiclassical state, the new state $\psi_{{\cal V}'}$ will not be one in general. 
To be precise, the semiclassicality conditions (for any choice of fiducial cell) tells us that $\sigma$ should be of the order of $\sqrt{2\alpha\, k_0}$. Thus, we can assume that $\psi_{\cal V}(k_0,p_0,\sigma)$ satisfies this condition. But then, for the new state  $\psi_{{\cal V}'}(k_0',p_0',\sigma')$ we have 
$$\sigma'=\ell\,\sigma=\sqrt{2\ell^2\alpha\,k_0}=\ell^{1/2}\,\sqrt{2\alpha k_0'}\, . 
$$
Thus, $\sigma'$ will satisfy the semiclassicality criteria for cell ${\cal V}'$ only if ${\ell}^{1/2}$ is not very different from one. Let us now consider the canonical mapping between the set of preferred semiclassical states, where $\sigma=\t\sigma(k_0)$, coming from the mapping on the labels defining the state, induced by the classical symmetry. In this case, even the expectation value of $\hat{p}_\phi$ will not be invariant, but instead,
\[
\la\hat{p}_\phi\ra'=\hbar k_0'\left(1 + \f{{\sigma'}^2}{4{k_0'}^2}\right)=\ell \hbar k_0\left(
1 +\f{1}{\ell} \f{{\sigma}^2}{4{k_0}^2}\right)\, .
\]
We can now see what is the relative difference with respect to the `rescaled' value $\ell\la\hat{p}_\phi\ra$ is then,
\be
\f{|\la\hat{p}_\phi\ra' - \ell \la\hat{p}_\phi\ra|}{\ell\la\hat{p}_\phi\ra}= \f{\alpha}{2 k_0} \left|1 -\f{1}{\ell}\right| + O(\alpha^2/k_0^2)\, .
\ee
Note then that, if the original state was semiclassical, so $k_0\gg \alpha$, then the relative difference will be small provided $\ell$ is of order one or larger. One can then say that there is approximate scale invariance. If, on the other hand, $\ell\ll 1$, then the relative difference can be large and therefore the scaling symmetry is not recovered even approximately.

Furthermore, we can also look at the expectation value of the density at the bounce Eq.(\ref{rho}), an observable that is cell independent (and therefore, `gauge independent'). It is straightforward to see that the terms of the form $\alpha^2/k_0^2$ and $\alpha^2/\sigma^2$ in that expression will break the `rescaling independence'. This can be seen from the expression (\ref{Vmin}) of the volume at the bounce that possesses such terms. One might, of course, consider more general class of states for which a satisfactory `cell re-scaling mapping' might be well
defined. We shall not pursue that avenue here.

We can conclude that only in the limit of large $k_0$, with respect to
$\alpha$, the density at the bounce approaches the critical density and that `cell independence' is approximately recovered for our class of semiclassical states. Let us now
introduce the weaker notion of {\it approximate rescaling invariance} for generic
observables. By this we mean the following. To begin with, we keep our semiclassicality conditions that, for each value of $k_0$, imply the choice $\sigma=\tilde{\sigma}(k_0)$, and then consider the behavior of expectation values and fluctuations. Then, we will say that
approximate rescaling invariance is implemented if the expectation values of `true observables' $\hat{\cal O}_i$, that classically are cell independent (and therefore of observational relevance), such as energy density, expansion and curvature, `converge' to the corresponding classical value for large $k_0$. Let us be precise. Given some $K>0$, and for any two values $k_0$ and $k'_0$ (both larger than $K$), there is convergence if the difference in the corresponding expectation values are smaller than any prescribed tolerance 
$\epsilon$: $|\la\hat{\cal O}_i\ra - \la\hat{\cal O}_i\ra'|<\epsilon$, for all $k_0,k_0'>K$. Similarly, we can also impose convergence conditions on the corresponding fluctuations of (cell independent) quantities. We can see from
the expressions above that in the solvable model of loop quantum cosmology, approximate rescaling invariance is
satisfied for our class of states. Thus, for a sufficiently large $K$ and appropriately chosen tolerance $\epsilon$, the observable predictions of the theories defined by cells ${\cal V}$ and ${\cal V}'$, as defined by semiclassical states satisfying our above conditions, are observationally indistinguishable. Finally,
note that scaling symmetry is a feature of the open $k$=0 model, and not of the particular quantization chosen. Thus, the issue is also present in the Wheeler-De Witt quantum theory for this model, and the discussion in that case would follow very closely what we have here considered.

\section{Squeezed States}

Let us now take the initial states as
\be \label{sq-state}
\t F(k)=\left\{
\begin{array}{rl}
k^n e^{-\eta(k-\beta)^2},& \mbox{ for } k>0 , \,\,\, \eta ,\beta \in {\mathbb C}\\[1ex]
0,& \mbox{ for } k\le 0 ,
\end{array}
\right.
\ee
depending on two complex parameters $(\eta,\beta)$.
One can reduce to a Gaussian state from a squeezed state by setting:
$
\mb{I}{\rm m}(\eta) =:\eta_I =0, \;\mb{R}{\rm e}(\eta)=: \eta_R=\f{1}{\sigma^2}, \;
\mb{R}{\rm e}(\beta) =: \beta_R=k_0, \; 2\eta_R\beta_I=-p_0 $.\\

\begin{figure}[tbh!]
\begin{center}
\includegraphics[scale=0.26]{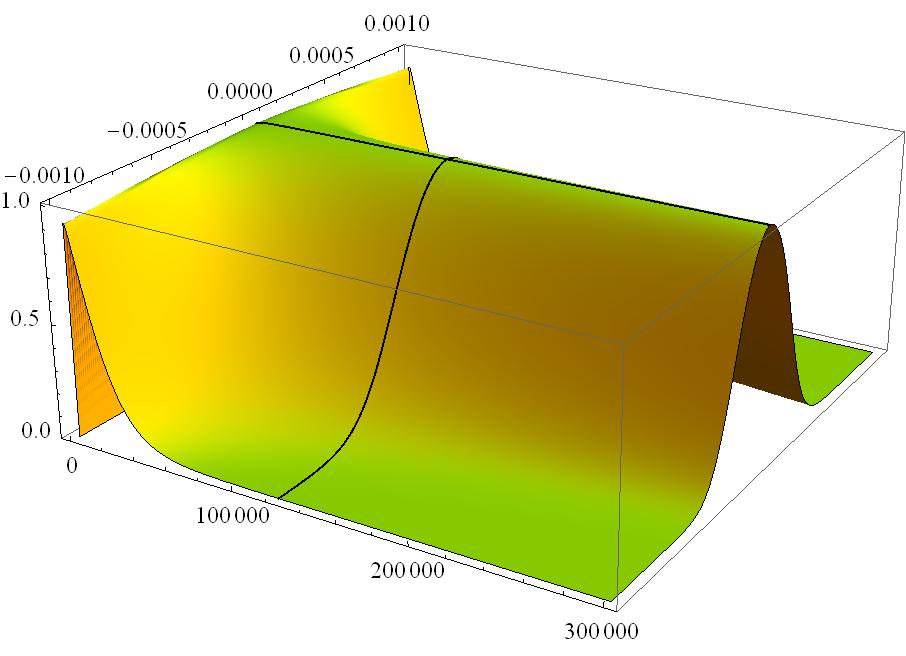}
\includegraphics[scale=0.26]{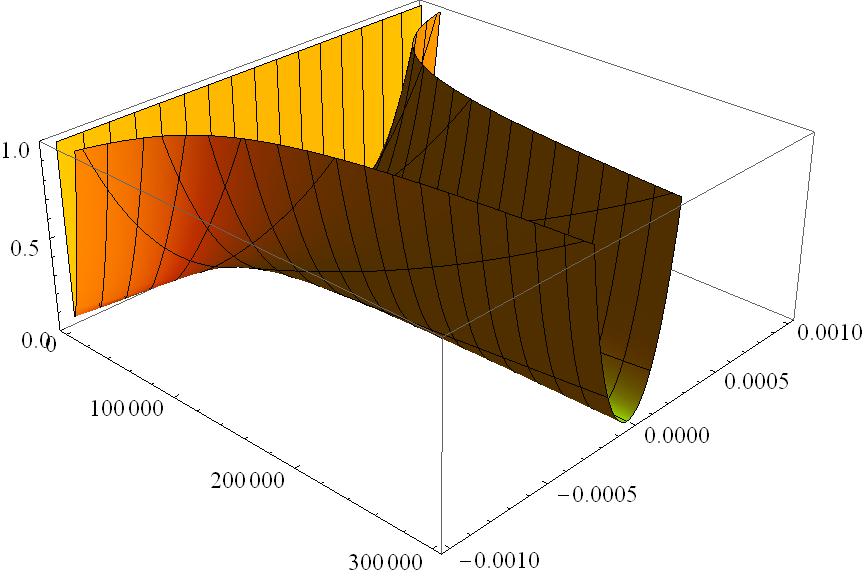}
\caption{Density (left) and asymptotic relative fluctuation $\Delta_+$ (right), as
functions of $\eta_I$ and $\sigma^2$.
		In the plot $1/\eta_R=\sigma^2\in (0, 300000)$,
		$\eta_I\in (-0.001,0.001)$ , for $\beta_R=k_0=10000$, $\beta_I=0$ and $n=0$.
        The maximal value of the density at the bounce correspond to the Gaussian states. 
        It is clear that the squeezed states have a maximal
        density close to $\rho_{\rm crit}$ only if the state is near to the black line corresponding to the $\eta_I=0$ Gaussian states. For the asymptotic relative dispersion the minimum is achieved close to the Gaussian states, and the asymptotic relative dispersion grows as one introduces squeezing.
}\label{fig:3}
\end{center}
\end{figure}
\begin{figure}[tbh!]
\begin{center}
\includegraphics[scale=0.4]{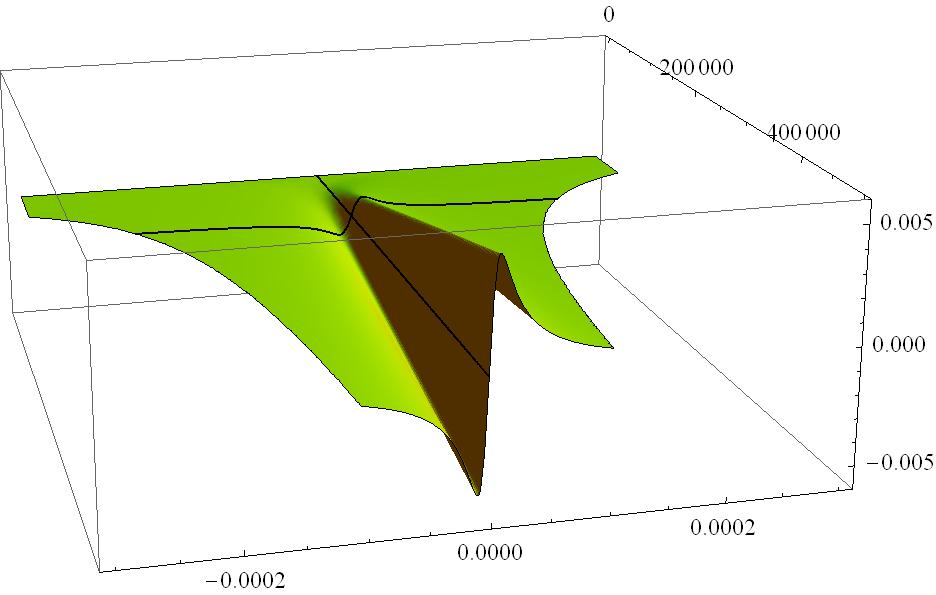}
\caption{We have plotted $\frac{\Delta_+-\Delta_-}{\Delta_-}$, the relative error 
of the difference in the asymptotic values of the relative fluctuation. 
		In the plot $1/\eta_R=\sigma^2\in (0, 10^6)$,
		$\eta_I\in (-0.0003,0.0003)$ , $\beta_R=k_0=10000$, $\beta_I=0$ and $n=0$.
        We have only plotted the surface in the region where
        the squeezed state have an asymptotic relative fluctuations 
        ($\Delta_+$ and $\Delta_-$) smaller than 1.
        The diagonal line marks the Gaussian states ($ \eta_I=0$) that is just where $\Delta_+ =   
        \Delta_-$, and the other line is for states with $\sigma=\t\sigma^2=2\alpha k_0$.
}\label{fig:4}
\end{center}
\end{figure}

Thus, we see that the extra parameter in the definition of the states is given by the imaginary
part of $\eta$.
How do squeezed states compare to Gaussian states? Are they any better?
Let us illustrate the answer to these questions in Figs.~\ref{fig:3} and \ref{fig:4}.
What these figures show is that we can go off the Gaussian states, by introducing the
{\it squeezing parameter} $\eta_I$, and the state will also have most of its main properties.
Semiclassicality is not destroyed, but the allowed range for the parameter $\eta_I$ is rather
small, so one has to remain very close to the Gaussian states. For details see \cite{CM1}.

For squeezed states the minimum of $\Delta_\pm$ (for constant $k_0$) is not on the Gaussian states $\eta_I=0$ but rather, for some value $\pm \t\eta_I$. If $\Delta_+$ reaches its minimum at $\tilde\eta_I$ then $\Delta_-$ does it for
$-\t\eta_I$. Thus, there is an asymmetry in the volume fluctuations. Can this feature destroy semiclassicality as sometimes is claimed? The short answer is, no.

For a semiclassical state, the relative change in $\Delta_\pm$ is very small (See Fig.~\ref{fig:4}). 
For instance, for $k_0=10^5$, the maximum of 
$|\Delta_+ -\Delta_-|/\Delta_\pm$ is about $10^{-4}$. As we increase $k_0$ this value decreases even further, inversely proportional to $k_0$. 
These results show that indeed, semiclassical states are very symmetric and that {\it semiclassicality is preserved across the bounce}. 
One should also note that this quantitative analysis invalidates some claims  regarding the allowed asymmetry in fluctuations across the bounce and `cosmic forgetfulness' \cite{odc,harmonic}, supports the results of \cite{cs:prl} and \cite{kam:paw}, and considerably improves the bounds found in \cite{kam:paw}.

Let us now see how our results
compare to those of  \cite{kam:paw}. In there, the authors prove, using general scattering methods not tailored to a specific choice of states, that for states that are sharply peaked on one side of the bounce the following relation is satisfied,
\be
|\sigma_+ - \sigma_-| \leq 2\sigma_*\, ,\label{KP-bound}
\ee
where $\sigma_\pm$ is the dispersion of the logarithm of $\hat{V}$ asymptotically after and before the bounce, and 
$\sigma_*$ is the dispersion of the logarithm of $\hat{p}_\phi$. If a state is sharply peaked such that the dispersions are small, one can safely assume that, for an observable $\hat{O}$,
$\langle\ln(\hat{O})\rangle\approx \ln\langle\hat{O}\rangle$ and $\langle \Delta \ln(\hat{O})\rangle \approx \langle\Delta\hat{O}\rangle/\langle\hat{O}\rangle$. With these assumptions, the
bound (\ref{KP-bound}) is well approximated by the following bound: $|\Delta^{1/2}_+ -\Delta^{1/2}_-|\leq 2 (\Delta\hat{p}_\phi)/\langle\hat{p}_\phi\rangle$. We can now exactly evaluate the quantities in this last expression to see whether the bound (\ref{KP-bound}) is close to being saturated for our semiclassical squeezed states. What we did was to start with the
squeezed states with the optimal $\sigma=\t\sigma$, and find the maximum value of the quantity
\be
{\cal E} := \f{|\Delta^{1/2}_+ -\Delta^{1/2}_-|}{2 ((\Delta\hat{p}_\phi)/\langle\hat{p}_\phi\rangle)}
\ee
for all allowed values of $\eta_I$, and taking $\eta_R=1/\tilde{\sigma}^2$, $\beta_I=0$. We have also taken $\beta_R=k_0$ large. The question is whether ${\cal E}$ is close to one for these states. What we saw is that this quantity ${\cal E}$ is not only much smaller than one for realistic values of the parameters, but it also decreases faster than  $C/k_0$. That is, ${\cal E}$
behaves as  ${\cal E}< C/k_0$, with $C\approx 3$. Thus, for realistic values for $k_0$, of the order of $k_0\approx 10^{120}$ corresponding to a `large universe' --in the spatially compact case-- the bound is very far from being saturated.  The bound (\ref{KP-bound}) is a very general result that hold for a large class of states, and ensures the preservation of semiclassicality. What we have shown here is that a much stronger bound is satisfied by our special class of squeezed semiclassical states. 

\section{Discussion}

The main objective of this article was to explore the semiclassical sector of loop quantum cosmology. That is, whether one could find explicit semiclassical states sharply peaked around a classical configuration. With this
objective in mind,
we saw that the solvable model in loop quantum cosmology is the perfect arena for asking concrete analytical questions regarding the semiclassical limit, and has allowed us to gain some conceptual clarity on the
issue.
As we have shown, coherent generalized Gaussian states are enough to construct well behaved, semiclassical states that are `peaked' on a classical configuration. Furthermore, we have seen that the choice of parameters that characterize the `spreading' of the wave function is severely restricted by physical considerations. Still, one can define `many' coherent states that are semiclassical, and represent an over-complete set of states. Even when 
we have benefited from the existence of the solvable model, one expects that there exist corresponding semiclassical states for other cosmological models, such as $k$=$\pm 1$ FRW and a cosmological constant, for example.
Many of the conceptual issues that we have encountered here will be common to those models as well.

We have further studied the behavior of these semiclassical states at the bounce, and have found that they maintain their coherence, having bounded fluctuations.  This has also allowed us to put the so called `effective theory' to the test. As we have seen, one can intuitively label states as more `semiclassical', when the point on phase space
where they are peaked represents a trajectory yielding a {\it large} universe\footnote{in the sense that the
volume at the bounce is large in Planck units.}.
Using this notion, we have seen that the more semiclassical the state is, the better the effective description, even at the bounce.
A somewhat important question one should ask is that of the {\it scaling symmetry} present in the classical and effective descriptions. What is then the situation of this symmetry in the quantum theory? As we have seen, 
one can not define a mapping, within our family of coherent states,
that has the desired transformation properties in terms of expectation values of operators, {\it and} remains
semiclassical for arbitrary cell rescaling. Nevertheless, we have shown that in the limit in which one considers
{\it large} universes, one recovers precisely the classical scaling symmetry, as one might have expected from simple considerations.


Finally, we have explored generalized  squeezed states and compared them to the generalized Gaussian states. As before, even when the region in parameter space where the states behave semiclassical is small, there is still a large number of states satisfying our conditions. We have also seen that squeezed states, that can be seen as an extension of Gaussian states, do not represent an improvement over the (particular case) of Gaussian states.
As we have argued, the bounce is very nearly symmetric regarding fluctuations of relevant operators, 
for all the allowed squeezed states.
Fluctuations are bounded before the {\it big bounce} if the state is semiclassical on the other side. There is no {\it cosmic forgetfulness.}

\vskip0.5cm
\noindent
We thank A. Ashtekar, T. Pawlowski and P. Singh for discussions and comments. 
This work was in part supported by DGAPA-UNAM IN103610 grant, by NSF
PHY0854743 and by the Eberly Research Funds of Penn State.

\end{document}